\documentclass[aps, 10pt, amsmath, twocolumn, superscriptaddress]{revtex4-1}
\usepackage[plainpages=false,colorlinks=true,linkcolor=blue, citecolor=blue, urlcolor=blue]{hyperref}
\usepackage[english]{babel}
\usepackage{graphicx}
\usepackage[charter]{mathdesign}
\usepackage{bm}
\usepackage[svgnames]{xcolor}
\usepackage{subfigure}
\usepackage{booktabs}

\usepackage{array}

\newcolumntype{C}{>{$\displaystyle}c<{$}}
\newcolumntype{L}{>{$\displaystyle}l<{$}}
\newcolumntype{R}{>{$\displaystyle}r<{$}}
\newcolumntype{A}{ >{$} r <{$} @{\extracolsep{0pt}} >{${}} l <{$} }

\newcommand\rv{\mathbf{r}}

\begin{document}

\title{Inter-layer bias effect on time-reversal symmetry breaking in twisted bilayer cuprates}

\author{Mathieu B\'elanger}
\affiliation{D\'epartement de physique and Institut quantique, Universit\'e de Sherbrooke, Sherbrooke, Qu\'ebec, Canada J1K 2R1}
\author{David S\'en\'echal}
\affiliation{D\'epartement de physique and Institut quantique, Universit\'e de Sherbrooke, Sherbrooke, Qu\'ebec, Canada J1K 2R1} 
\date{\today}

\begin{abstract}
We study a one-band Hubbard model of twisted bilayer cuprates with a twist angle of $53.13^\circ$. By introducing an inter-layer bias, we simulate heterobilayers of different dopings. Using the variational cluster approximation (VCA) we probe the effect of this bias on the time-reversal-symmetry (TRS) breaking phase. Doping differences between layers affect the region where TRS-breaking occurs; we construct a phase diagram mapping out the TRS-breaking phase in the $n_1$-$n_2$ plane, $n_\ell$ being the electron density on layer $\ell$. We also map the spontaneous supercurrent on the same plane.
\end{abstract}
\maketitle

\section{Introduction}

Previous studies of the Hubbard model for twisted cuprates using the variationnal cluster approximation (VCA) at $\theta=53.13^\circ$~\cite{lu_doping_2022} and $43.60^\circ$~\cite{belanger_doping_2023} have shown that time-reversal symmetry (TRS) is spontaneously broken in a narrow region of the superconducting dome  when strong inter-layer tunneling is considered.
This shows that the TRS-breaking phase predicted around $45^\circ$~\cite{can_high-temperature_2021,volkov_magic_2023,song_doping_2022,fidrysiak_tuning_2023} is strongly doping dependent. It was proposed that such a phase may lead to Majorana modes when in proximity with a material with spin-orbit coupling~\cite{margalit_chiral_2022,mercado_high-temperature_2022,li_high-temperature_2023}.
Superconducting qubits using twisted cuprates have been proposed \cite{brosco_superconducting_2023}.
Extensions to multilayer systems have also been studied~\cite{tummuru_twisted_2022}.

The realization of two-dimensional monolayers of Bi${}_2$Sr${}_2$CaCu${}_2$O${}_{8+\delta}$ (Bi2212) with a transition temperature close to that of bulk samples~\cite{yu_high-temperature_2019,zhao_sign-reversing_2019} allows cuprate bilayers to be assembled in the laboratory and $c$-axis Josephson junctions to be created. Because of the $d$-wave pairing symmetry in each layer, the critical current changes depending on twist angle in those junctions~\cite{lee_twisted_2021,martini_twisted_2023,lee_encapsulating_2023,zhao_emergent_2021}. The critical current can remain finite at $45^\circ$,  pointing to the predicted TRS-breaking phase~\cite{volkov_josephson_2021,tummuru_josephson_2022}. 

Those junctions are challenging to make due to disorder inherent to Bi2212. It can thus be difficult for the two monolayers to be locally at the same doping. Indeed, the distribution of dopants can be inhomogeneous, or the preparation process can introduce defects. On the other hand, it was proposed that some inhomogeneity could be needed in order to induce TRS breaking in twisted cuprate junctions~\cite{yuan_inhomogeneity-induced_2023}.

Since changing the doping results effectively in a different material, one can take inspiration from the heterobilayer transition metal dichalcogenides~\cite{ruiz-tijerina_interlayer_2019,tang_simulation_2020} and use different monolayer cuprates to create the bilayer system. 
The physics of cuprates being doping dependent, this would affect the TRS-breaking phase.

The Hubbard model used in Refs.~\cite{lu_doping_2022,belanger_doping_2023} can be modified to introduce a doping difference between the layers. 
This can also simulate the effect of defects or contamination in the junction-making process leading to close, but different, doping content in each layer.

In this paper we introduce an inter-layer bias in the twisted cuprates Hubbard model at $\theta=53.13^\circ$ studied in Ref.~\cite{lu_doping_2022}. This bias induces a doping difference between the two layers, allowing us to simulate heterobilayers cuprates. We show that the doping range where TRS breaking occurs is affected by inter-layer bias. We obtain a phase diagram mapping out the TRS-breaking phase in the $n_1$-$n_2$ plane, $n_\ell$ being the electron density on layer $\ell$. 
We also compute the spontaneous supercurrent circulating in a certain loop within the TRS-breaking phase; this can be used as a TRS-breaking order parameter.

\section{Model}
\label{sec:model}

We use the Hamiltonian proposed in Ref.~\cite{lu_doping_2022}, where each layer is described by a one-band Hubbard model, each site corresponding to a copper atom. 
To this layer Hamiltonian we add a inter-layer bias term $H_\epsilon$, so that the complete Hamiltonian is
\begin{align}
H=H^{(1)}+H^{(2)}+H_\perp+H_\epsilon,
\label{eq:H}
\end{align}
where the intra-layer Hamiltonian $H^{(\ell)}$ is
\begin{align}
H^{(\ell)}=\sum_{\rv,\rv'\in \ell,\sigma}t_{\rv \rv'}c^\dagger_{\rv,\ell,\sigma}c_{\rv',\ell,\sigma}+U\sum_{\rv }n_{\rv,\ell,\uparrow}n_{\rv,\ell,\downarrow}-\mu\sum_{\rv,\sigma}n_{\rv,\ell,\sigma}~.
\label{eq:H-intra}
\end{align}
$c_{\rv,\ell,\sigma}$ $(c^\dagger_{\rv,\ell,\sigma})$ is the annihilation (creation) operator of an electron at site $\rv $ on layer $\ell$ with spin $\sigma=\uparrow,\downarrow$, and $n_{\rv,\ell,\sigma}$ is the number operator. 
$\rv,\rv'$ are the site indices of a square lattice for each layer. 
The on-site repulsion between electrons is $U$. 
The hopping matrix $t_{\rv \rv'}$ includes nearest-neighbor hopping ($t$) and next-nearest-neighbor hopping ($t'$). 
To describe Bi2212 we use the values $t=1$, $t'=-0.3$ and $U=8$ with $t$ being the energy unit~\cite{markiewicz_one-band_2005,lu_doping_2022,belanger_doping_2023}.
Nonlocal interactions were not considered since superconductivity can be driven by local repulsion alone and is resilient to nearest-neighbor repulsion at intermediate to strong coupling~\cite{senechal_resilience_2013}.

The coupling between the layers is provided by inter-layer tunneling:
\begin{align}
H_\perp=\sum_{n=1}^{3}V_n\sum_{\langle\rv,\rv'\rangle_{\perp,n,\sigma}}\left[c^\dagger_{\rv,1,\sigma}c_{\rv',2,\sigma}+\mathrm{H.c.}\right],
\label{eq:H-inter}
\end{align}
with $\langle\rv,\rv'\rangle_{\perp,n,\sigma}$ representing the set of sites $\rv $ on layer 1 and $\rv'$ on layer 2, such that their projections on the plane are $n^{\rm th}$ neighbors. 
We consider inter-layer hopping up to third inter-layer neighbors. 
The strength of the tunneling is given as in Ref.~\cite{belanger_doping_2023} by 
\begin{align}
V_n=Ve^{-\lambda\left(|\mathbf{d_n}|-d_z\right)/a}, 
\end{align} 
where $|\mathbf{d_n}|=|\rv -\rv'|$ is the three-dimensional distance between the two sites corresponding to the $n^{\rm th}$ neighbors on different layers, $d_z$ is the distance between the two layers and $a$ is the lattice constant of the square lattice. 
The inter-layer tunneling between sites that are on top of each other is $V$. 
We use a damping parameter $\lambda$, the same as in Ref.\cite{belanger_doping_2023}: $d_z=a$ and $\lambda=11.13$. 
This set of parameters leads to similar inter-layer tunneling as in Ref.~\cite{lu_doping_2022}. 
We use $V=0.4$, since a strong inter-layer tunneling is needed to have a clear TRS breaking~\cite{lu_doping_2022, belanger_doping_2023} .

The inter-layer bias term takes the form
\begin{align}
H_\epsilon=-\epsilon\sum_{\rv,\sigma} \left( n_{\rv,1,\sigma}-n_{\rv,2,\sigma}\right)~.
\end{align}
This contribution effectively shifts the chemical potential on each layer by $\pm\epsilon$. 
The density $n_\ell$ in each layer is then different from the total density $n$. The transformation $\epsilon\to-\epsilon$ effectively swaps both layers so we can concentrate on positive values of $\epsilon$. 

The two layers are assumed to have the same lattice constant. Different lattice constants would not lead to a commensurate unit cell with a reasonable number of orbitals with twist angle close to $45^\circ$.  

\begin{figure}[h]
\begin{center}
\includegraphics[width=0.9\hsize]{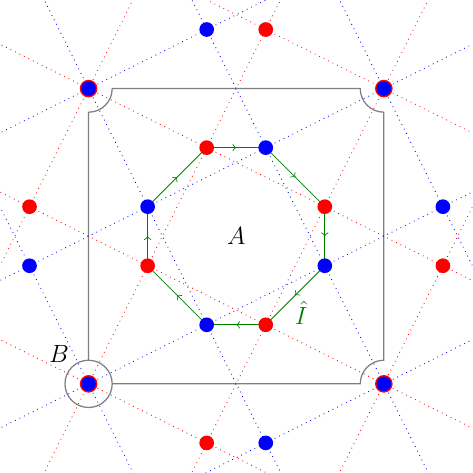}
\caption{Unit cell of the twisted bilayer cuprate system at $\theta=53.13^\circ$, containing 10 sites between the two layers.  The top (bottom) layer correspond to the blue (red) lattice. The $A$ clusters contain 8 sites and the $B$ cluster contains only 2 sites that are on top of each other. The green arrow show the direction of the current defined by Eq.~\eqref{eq:current}.
\label{fig:schema_10sites}}
\end{center}
\end{figure}

Model~\eqref{eq:H} is applied to the bilayer with twist angle $\theta=53.13^\circ$. At this twist angle the unit cell of the bilayer system is made of 10 sites, as illustrated on Fig.~\ref{fig:schema_10sites}.
That twist angle was chosen over $43.60^\circ$ because of the relatively low computing resources needed. 

The superconducting phase in this model is probed using the VCA~\cite{potthoff_variational_2003,dahnken_variational_2004} with an exact diagonalization solver at zero temperature, like in Refs~\cite{lu_doping_2022,belanger_doping_2023}. 
This variational method on the electron self-energy, based of Potthoff's self-energy functional approach, allows us to probe broken symmetries while preserving strong correlations. 
It has been used to study magnetic phases~\cite{dahnken_variational_2004,sahebsara_hubbard_2008} and superconductivity~\cite{faye_interplay_2017,belanger_superconductivity_2022} in various systems. For a detailed review of the method, see Refs.~\cite{lu_doping_2022,potthoff_variational_2012,potthoff_cluster_2018}.

As shown in Ref.~\cite{lu_doping_2022}, we expect the superconducting order parameter of the bilayer system to belong to the irreducible representations $B_1$ or $B_2$ of the $D_4$ point group of the bilayer. We define the VCA Weiss field belonging to these two representations as
\begin{align}
\hat{B}_1=\hat{\Delta}^{(1)}+\hat{\Delta}^{(2)},\quad \hat{B}_2=\hat{\Delta}^{(1)}-\hat{\Delta}^{(2)},
\end{align}
where the $d$-wave pairing operator on layer $l$ is defined as
\begin{align}
\begin{split}
\hat{\Delta}^{(\ell)}=&\sum_{\rv \in \ell}c_{\rv,\ell,\uparrow}c_{\rv +\mathbf{x}^{(\ell)},\ell,\downarrow}-c_{\rv,\ell,\downarrow}c_{\rv +\mathbf{x}^{(\ell)},\ell,\uparrow}\\
&\quad- c_{\rv,\ell,\uparrow}c_{\rv +\mathbf{y}^{(\ell)},\ell,\downarrow}+c_{\rv,\ell,\downarrow}c_{\rv +\mathbf{y}^{(\ell)},\ell,\uparrow}.
\end{split}
\end{align}
For a more detailed description and justification of these definitions, see Refs~\cite{lu_doping_2022,belanger_doping_2023}.

In the VCA procedure, we can use $\hat{B}_1$ or $\hat{B}_2$ to probe the superconducting phase. One of them should lead to a lower-energy state and be favored. It is also possible that the complex combination $\hat{B}_1+i\hat{B}_2$ lowers the energy even more; this combination corresponds to the TRS-breaking state. In such cases we can express the relative phase $\phi$ between the order parameters $\langle\hat{\Delta}^{(1)}\rangle$ and $\langle\hat{\Delta}^{(2)}\rangle$ of the two planes as 
\begin{align}
\tan\frac{\phi}{2}=\frac{\text{Im} \psi_{B_2}}{\text{Re} \psi_{B_1}},
\label{eq:phase}
\end{align}
with $\psi_{B_i}$ the order parameter $\psi_{B_i}=\frac{1}{L}\langle \hat{B}_i\rangle$, where $L$ is the number of site and $i=1,2$. 
A value of $\phi=0$ ($\phi=\pi$) corresponds to a pure $B_1$ ($B_2$) case. The interesting case is the one where $\phi\neq 0$ or $\pi$, where there is a coexistence of both states, indicating a TRS breaking.

We use the VCA procedure with Weiss fields from both representations ($B_1$ and $B_2$), with varying values of $\epsilon$, to probe the effect of different layer doping content on the TRS-breaking phase.

\section{Result and discussion}
\label{sec:doping}
\subsection{inter-layer bias}

\begin{figure}[ht]
\begin{center}
\includegraphics[width=\hsize]{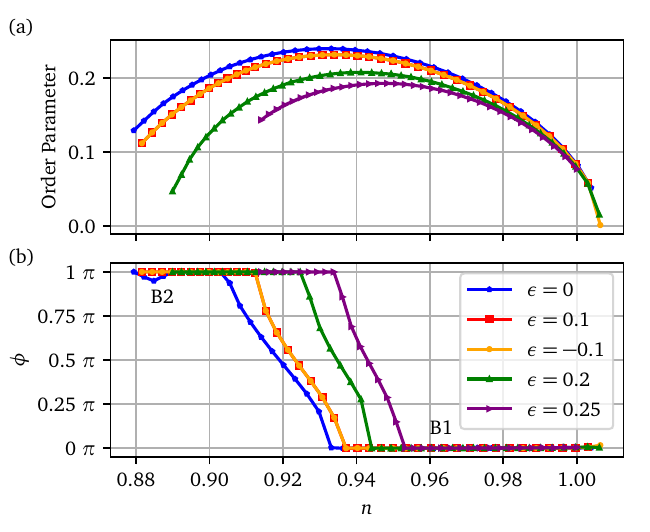}
\caption{Superconducting order parameter as a function of electron density $n$ in the twisted cuprate bilayer at $\theta=53.13^\circ$ for different inter-layer bias parameters $\epsilon$. (a) Norm $\sqrt{|\psi_1|^2+|\psi_2|^2}$ of the order parameter obtained from the VCA procedure with both representation as Weiss field for different values of $\epsilon$. The order parameter drops when $|\epsilon|$ increases. (b) Relative phase $\phi$ between the two layers. We observe a shift in the range of the TRS breaking close to half-filling ($n=1$) when $|\epsilon|$ increases. 
\label{fig:phase_ep}}
\end{center}
\end{figure}

Fig.~\ref{fig:phase_ep} shows the superconducting order parameter and relative phase $\phi$ as a function of electron density in model~\eqref{eq:H} with $\theta=53.13^\circ$, for different values of $\epsilon$. In Fig.~\ref{fig:phase_ep} (a), we observe that the difference in doping between the two layers causes a drop of the order parameter.  Assuming a monotonous relation between the order parameter and the critical temperature $T_c$, we can infer that $T_c$ should be maximal when $\epsilon=0$ (when both layer are identical). As expected from the symmetry of the system, the effect depends on the absolute value $|\epsilon|$ only. This can also be seen in Fig.~\ref{fig:phase_ep} (b), where we show the relative phase $\phi$. The TRS-breaking phase corresponds to the region where $\phi\neq0$ or $\pi$. We observe a shift in TRS-breaking region towards half-filling with increasing $|\epsilon|$.
While increasing $|\epsilon|$ the TRS doping range also becomes narrower, making it more difficult to detect at high values of bias $\epsilon$. If the doping discrepancy between the two layers is too large, the system may not show the TRS breaking behavior. This might explain the difficulty to observe a non-zero critical current in some $45^\circ$ $c$-axis Josephson junctions~\cite{lee_twisted_2021,martini_twisted_2023,lee_encapsulating_2023,zhao_emergent_2021}.

For all values of $\epsilon$ considered, the TRS breaking occurs in the overdoped region, i.e., beyond optimal doping according to Fig.~\ref{fig:phase_ep}, in at least one of the layers. 
There are theoretical signs that the superconducting states in the under- and overdoped regions are qualitatively different, even though they share the same symmetry~\cite{dash_pseudogap_2019}. 
Correlation effects being lower in the overdoped region, the superconducting state is closer to the BCS state than in the underdoped region. This seem to impact the TRS-breaking phase.

\begin{figure}[h]
\begin{center}
\includegraphics[width=\hsize]{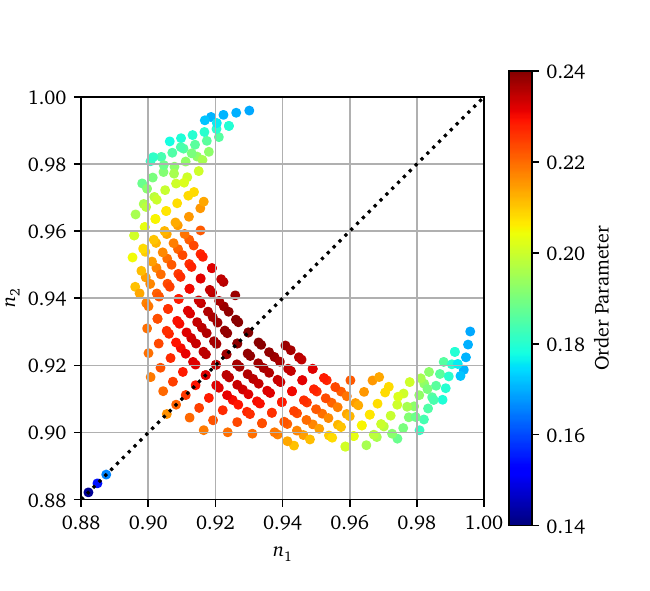}
\caption{ Phase diagram of the TRS breaking phase for different sets of layer doping $(n_1,n_2)$. The points indicate that a non-trivial relative phase was found with this combination of layer dopings. The color map represents the order parameter. The order parameter is maximum when the bias $|\epsilon|$ vanishes (dotted line).
\label{fig:doping_map}}
\end{center}
\end{figure}

In Fig.~\ref{fig:doping_map} we show a map of the TRS-breaking phase as a function of doping $n_{1,2}$ on each layer. 
The diagram has a crescent form and is symmetric around the zero bias ($\epsilon=0$) corresponding to $n=n_1=n_2$.
The order parameter drops when deviating from $\epsilon=0$, as seen in Fig.~\ref{fig:phase_ep}. 

The distribution of the TRS-breaking phase is not uniform. In fact, some combinations offer a bigger tolerance to doping differences. Indeed, when one of the layers is in the overdoped region $n_\ell\in [0.90,0.92]$, TRS-breaking occurs in a larger interval of doping for the second layer. On the other hand, near the tips of the crescent, the system has a small tolerance to doping difference and a TRS-breaking phase will be hard to observe.

The three points close to $n=0.88$ on the dotted line correspond to the small bump seen in Fig.~\ref{fig:phase_ep} for $\epsilon=0$. We believe that those results are an artefact of the method and do not hold physical meaning since no other value of $\epsilon$ exhibit this behavior.  
\begin{figure}[h]
\begin{center}
\includegraphics[width=\hsize]{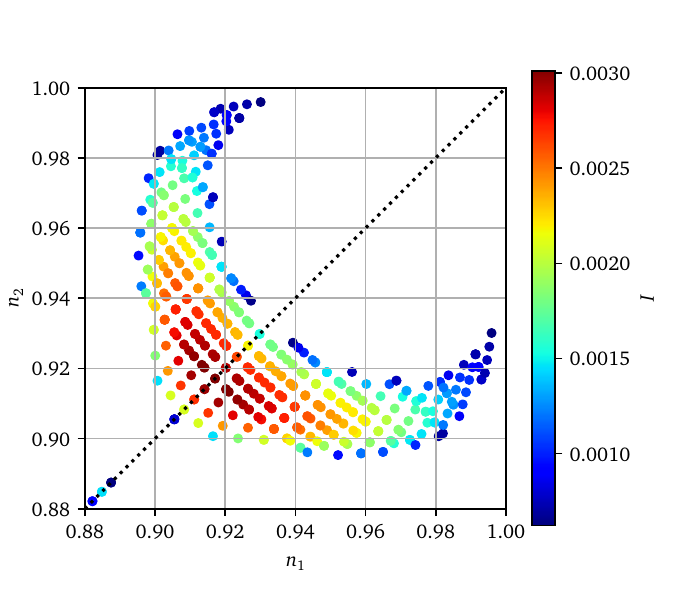}
\caption{Map of the spontaneous current $I$ along the loop defined in Fig.~\ref{fig:schema_10sites} for different sets of layer doping $(n_1,n_2)$. A dot indicates that a nonzero spontaneous current ($|I|>10^{-4}$) and the color map represents the value of the current. The dotted line corresponds to $\epsilon=0$.
\label{fig:I_map}}
\end{center}
\end{figure}

\subsection{inter-layer current}

It is possible to define a inter-layer current operator $\hat{I}$ between the sites of the different layers on cluster $A$ as
\begin{align}
\hat{I}= i\sum_{\{\rv,\rv'\}_I} \left( c_{\rv,1,\sigma}^\dagger c_{\rv',2,\sigma} -
c_{\rv',2,\sigma}^\dagger c_{\rv,1,\sigma}\right),
\label{eq:current}
\end{align}
where $\{\rv,\rv'\}_I$ is the set of pairs of sites defining the green path in Fig.~\ref{fig:schema_10sites}. This operator can be used to extract information related to a Josephson current, with the order parameter given by $I=\frac{1}{L}\langle \hat{I}\rangle$. Experimentally, a non-zero Josephson current appear when the relative phase between both layer is non-trivial. The maximal current correspond to a relative phase of $\phi=\frac{\pi}{2}$. This behavior is observed within our data while using Eq.~\eqref{eq:current} as the definition of our Josephson current.

Fig. \ref{fig:I_map} show the phase diagram for different sets of layer doping $(n_1,n_2)$. 
The points indicate that a current $|I|>10^{-4}$ was found, whose intensity is mapped in color.
This criterion makes sure that the current is significantly larger than the numerical precision ($10^{-7}$).
The current is maximum when $\phi$ is close to $\frac{\pi}{2}$ and when the two layers have similar doping levels.
The crescent has the same shape as in Fig.~\ref{fig:doping_map}, except that the current falls to zero outside of the crescent, whereas the SC order parameter does not. The current is indeed an order parameter for TRS-breaking.
The choice of current loop in Fig.~\ref{fig:schema_10sites} is not the only one possible. 
Other closed paths between the two layers would yield similar results, except for the overall current amplitude. 
Without external bias, we expect the net current between the two layers to vanish.

\subsection{Effect of $t'$}

The value of the next-nearest-neighbor hopping ($t'$) was chosen to best describe Bi2212. 
It is possible to change this value to probe the effect of considering different compounds. 
We looked at two other values of $t'$ ($-0.2$ and $-0.45$) while keeping every other parameter the same ($t=1$, $U=8$ and $V=0.4$). 

\begin{figure}[ht]
\begin{center}
\includegraphics[width=\hsize]{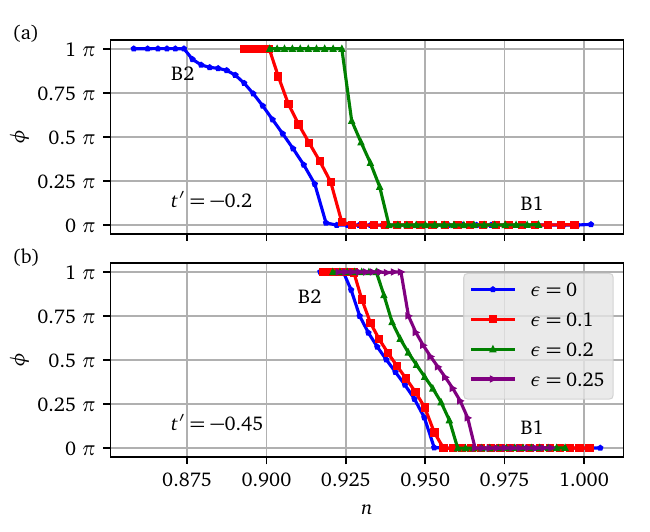}
\caption{(a) Relative phase $\phi$ between the two layers for $t'=-0.2$. The TRS-breaking  
region is shifted to higher doping compared to $t'=-0.3$ (Fig.~\ref{fig:phase_ep}). 
(b) Relative phase $\phi$ between the two layers for $t'=-0.45$. In this case, the TRS-breaking  
region is shifted to lower doping compared to to $t'=-0.3$.
\label{fig:phase_param}}
\end{center}
\end{figure}

Fig.~\ref{fig:phase_param} show the relative phase $\phi$ obtained by VCA for two other values of $t'$. 
The doping range where the TRS-breaking phase is observed is shifted when $t'$ is changed from $-0.3$. 
For $t'=-0.2$, the region is shifted toward higher doping, while for $t'=-0.45$ it is shifted toward half-filling. 
This shows that the TRS-breaking phase is robust against changes in the dispersion.

\begin{figure}[h]
\begin{center}
\includegraphics[width=\hsize]{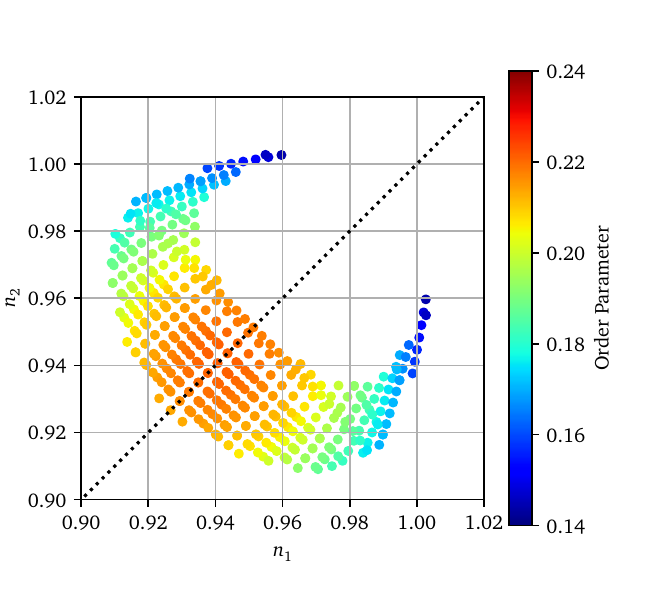}
\caption{Phase diagram of the TRS-breaking phase for different sets of layer density $(n_1,n_2)$ for $t'=-0.45$. The features observed here are similar to what is observed for $t'=-0.3$ (Fig. \ref{fig:doping_map}).
\label{fig:doping_map_0.45}}
\end{center}
\end{figure}

Fig.~\ref{fig:doping_map_0.45} show the phase diagram of the TRS-breaking phase for combinations of layer density $(n_1,n_2)$ for $t'=-0.45$. 
The shape of the diagram is similar to that for $t'=-0.3$, but shifted closer to half-filling. 
Some data points show a density $n_\ell>1$, this can be attributed to the error on the electron density typical of VCA when the chemical potential within the cluster is not treated as an additional variational parameter.

From the results presented here for Model~\eqref{eq:H}, it is possible to explain the sensibility of the cuprate Josephson junction to impurities and doping. 
At the same time, if one layer is in the high-tolerance region, the TRS-breaking phase could be easier to obtain. 
We note that our model is an oversimplification of the cuprate bilayer since it is based on the one-band Hubbard model and ignores the fact that each layer of the twisted system is in fact itself a bilayer. Still, we hope that the effect of doping asymmetry presented here are robust.

\section{Conclusion}
\label{sec:conclusion}

We used a one-band Hubbard model describing twisted bilayer cuprates at $\theta=53.13^\circ$ with an inter-layer bias, $\epsilon$, simulating a doping asymmetry between layers. 
Using the variational cluster approach, we probed the superconducting phase and found that $|\epsilon|$ affects the doping range and order parameter of the time-reversal-breaking state. 
We use the spontaneous current along a small loop as a TRS-breaking order parameter.
Increasing the inter-layer bias pushes the TRS-breaking region towards half-filling while making it narrower. 
The SC order parameter also decreases when the inter-layer bias increases. 
Overall, the TRS region has a crescent shape in the $n_1$-$n_2$ plane ($n_{1,2}$ being the electron densities on layers 1 and 2). 
One of the layers has to be in the overdoped region for the bilayer to break time-reversal.
But once a layer is overdoped, there is some tolerance to a doping difference with the other layer. 


\begin{acknowledgments}
This work was supported by the Natural Sciences and Engineering Research Council of Canada (NSERC) under grant RGPIN-2020-05060, by the NSERC postgraduate scholarships doctoral program and by the {\it Fonds de Recherche du Qu\'ebec Nature et technologies} (FRQNT) doctoral research scholarships.
Computational resources were provided by the Digital Research Alliance of Canada and Calcul Qu\'ebec.
\end{acknowledgments}


%

\end{document}